\documentclass[reprint,amsmath,amssymb,aps,pra,citeautoscript,longbibliography,superscriptaddress]{revtex4-1}
\usepackage{graphicx}
\usepackage{dcolumn}
\usepackage{bm}
\usepackage{ulem,color}
\usepackage{hyperref}
	\hypersetup{colorlinks=true,linkcolor=blue,citecolor=red,urlcolor=blue}
\usepackage{physics}
\usepackage{color}

\begin{document}
\title{Quantum carpets from Gaussian sum theory}
	\author{Huixin Xiong}
		\affiliation{Department of Physics, Peking University, Beijing 100871, China}
			\author{Xue-Ke Song}
		\email{songxk@sustc.edu.cn}
		\affiliation{Institute for Quantum Science and Engineering and Department of Physics, Southern University of Science and Technology, Shenzhen 518055, China}
		\affiliation{Department of Physics, Southeast University, Nanjing 211189, China}
	\author{H. Y. Yuan}
		\affiliation{Institute for Quantum Science and Engineering and Department of Physics, Southern University of Science and Technology, Shenzhen 518055, China}
\author{Dapeng Yu}
\affiliation{Institute for Quantum Science and Engineering and Department of Physics, Southern University of Science and Technology, Shenzhen 518055, China}
\affiliation{Shenzhen Key Laboratory of Quantum Science and Engineering, Southern University of Science and Technology, Shenzhen, 518055, China}
\author{Man-Hong Yung}
\email{yung@sustc.edu.cn}
\affiliation{Institute for Quantum Science and Engineering and Department of Physics, Southern University of Science and Technology, Shenzhen 518055, China}
\affiliation{Shenzhen Key Laboratory of Quantum Science and Engineering, Southern University of Science and Technology, Shenzhen, 518055, China}

	\date{\today}
	\begin{abstract}
			In many closed quantum systems, an interesting phenomenon, called quantum carpet, can be observed, where the evolution of wave function exhibits a carpet-like pattern. The mechanism of quantum carpet is similar to the classical interference pattern of light. Although the origin of quantum carpets has been studied previously, there are still many interesting details worth exploring. Here, we presented a unified framework for a simultaneous analyzing on three features of quantum carpets, namely full revival, fractional revival and diagonal canal. For fractional revival, a complete formula is presented to explain its generation through ``Gaussian sum theory'', in which all the essential features, including the phases and amplitudes, of this phenomenon could be captured analytically. Moreover, we also revealed the relations between the interference terms of the diagonal canals and their geometric interpretations, providing a better understanding in the formation of diagonal canals.
	\end{abstract}
	\maketitle
\section{introduction}
	\par
		In quantum mechanics, the infinite-square-well model (also known as the ``particle-in-a-box'' model) remains one of the best physical models for illustrating various fundamental concepts in the quantum theory \cite{bonneau2001, robinett2004, fojona2010, waegell2016}. It captures the essential features of the bound-state problems with a deep confining potential, providing the first approximation to semiconductor quantum wells~\cite{bastard1985, polkovnikov1998, kotova2018}. All the eigenvalues and eigenvectors of the infinite-square-well model can be obtained analytically, and the dynamics can be solved through the method of eigenstate expansion~\cite{schif, griffiths,yuan2018,yuan2018theo}. Moreover, the dynamics of the infinite potential well with moving walls can be speeded up by shortcuts to adiabaticity~\cite{chen2010,campo,songnjp,songpra}. Even though the dynamics of the infinite-square-well model is well understood, its study continues inspiring new insights in the field of quantum physics \cite{belloni, lin2016, lin2017}.
	\par
		In particular, in the solution of the dynamics for a particle trapped inside a infinite-square-well potential, there exists an art-like pattern, so called ``quantum carpet'', which was introduced by Kaplan et al.~\cite{kaplan2000}. Quantum carpets depict the behavior of the wave function evolution in the space-time, exhibiting a key feature showing the emergence of the initial wave function after the``revival time''~\cite{berry2001quantum}.  Essentially, the occurrence of the revival patterns in quantum carpets are due to the phase alignment of the neighboring eigenstates during the evolution~\cite{nest2006, hornberger2012, munoz2016, barros2017,yousaf2016quantum, kazemi2013quantum,garcia2014, banchi2015, rohith2015, krizanac2016}. Experimentally, the properties of quantum carpets have been illustrated in a wide class of systems, for example,  Rydberg atom~\cite{yeazell1990observation}, optical lattice~\cite{greiner2002collapse}, cavity quantum electrodynamic systems~\cite{rempe1987observation}, optical waveguide~\cite{valle2009}, and so on.
	\par
		Apart from the full revival of the initial wave function, the time evolution in a quantum carpet also contains solutions cloning initial wave functions throughout the well --- a phenomena called ``fractional revival"~\cite{aronstein1997fractional,naqvi2001fractional,spanner2004,bernard2017graph}. It originates from the the phase alignment with {\it non-adjacent} eigenstates. Fractional revival represents a partial revival of wave function when the evolution is fractional multiples of the revival time \cite{dooley2014,rohith2014,lemay2016,christandl2017}. Different approaches have been developed in analyzing fractional revivals. For example, a previous analysis~\cite{aronstein1997fractional} on fractional revival  was based on the dynamics of classical-like wave packets, where the amplitudes and relative phases of fractional revival were obtained. Alternatively, one may apply Fourier analysis directly to analyze the fractional revival~\cite{naqvi2001fractional}, where absolute phases of fractional revival were obtained without invoking classical physics. However, the authors did not provide a complete account of fractional revival, in the sense that their results did not explain why the amplitudes of the copies are identical.
\par
Our goal here is to present a unified framework that is capable of recovering both the revival amplitudes and {\it absolute} phases in a fractional revival, without relying on classical physics. To this end, we show an exact formula of continuous quantum carpets to explain the fractional revival by ``Gaussian sum theory'', in which both the revival amplitudes and {\it absolute} phases can be analyzed in the same framework.
On the other hand, there exist other patterns of the continuous quantum carpet, diagonal canals, whose formation can be explained by decomposing the carpet into background and interference terms assisted by the Wigner function \cite{stifter1997teilchen,friesch2000quantum}. We shall also provide a detailed analysis about the relations between background/interference terms and their geometrical correspondences in the diagonal canals. Furthermore, we show the theoretical description of the discrete carpet, and then discuss its experimental application by optical waveguide.
	\par
		The paper is organized as follows: In Sec \ref{l002}, we show the analysis of general framework of continuous carpet, including full revival, fractional revival, and diagonal canal, and present the new formula for fractional revival and the explicit analysis of background/interference terms for the diagonal canal. In Sec \ref{l003}, we give the theoretical study of the discrete carpet at first, and then discuss its experimental implementation in optical waveguide. A summary is enclosed in Sec \ref{l004}.
\section{continuous quantum carpets} \label{l002}
	\subsection{Full revival of wave function}
		\par
\begin{figure}[htbp]
			\centerline{\includegraphics[width=\linewidth]{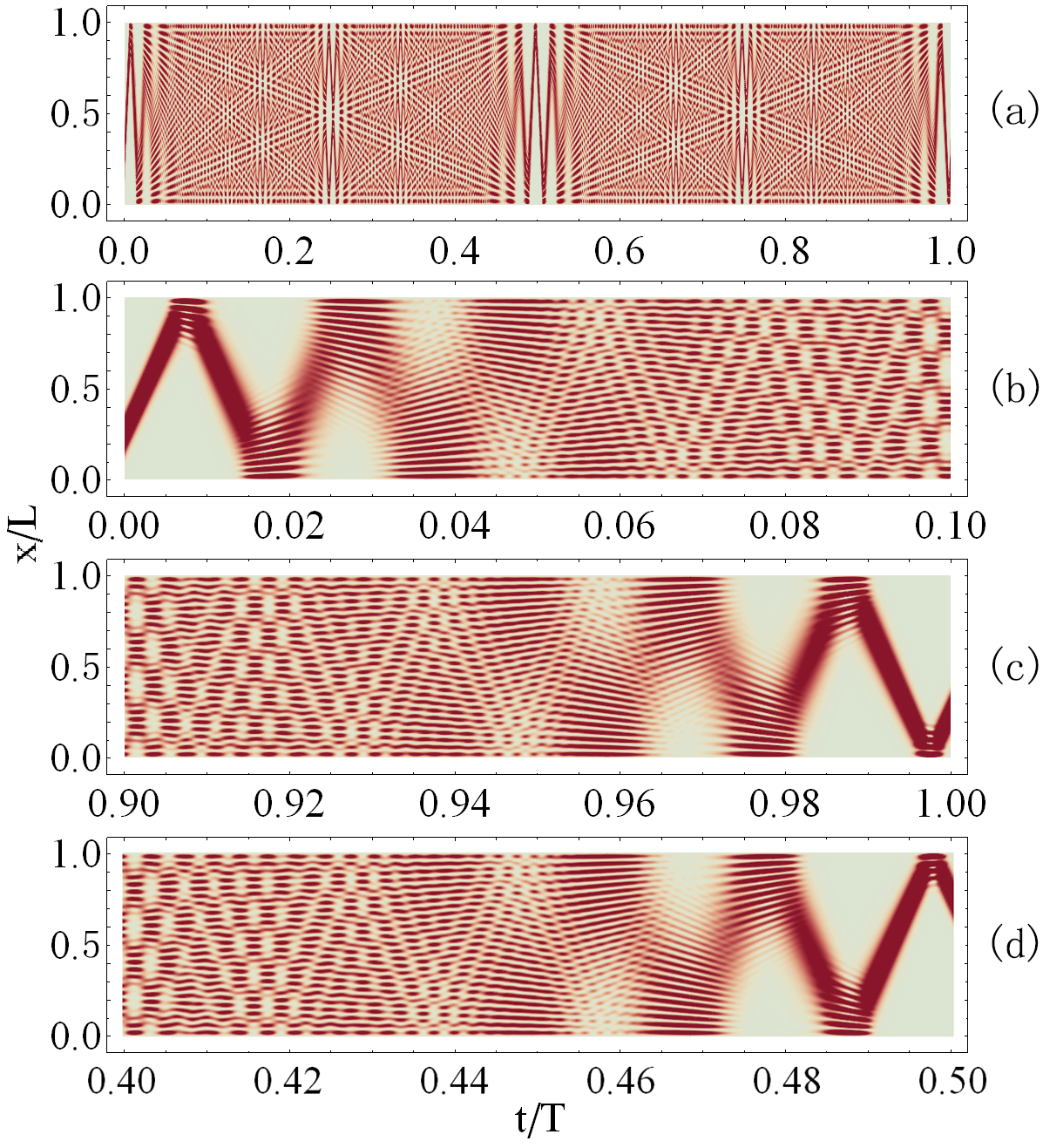}}
			\caption{Full and mirror revival of the quantum carpet.	$x$ is the position of the particle, $L$ is the width of the well, $t$ is the evolution time, and $T$ is the revival time. Parameters $\bar x / L = 1/4$, $s_x/L=1/\left(5\pi\right)$, $\bar p/\left(\hbar/L\right)=25\pi$.	The probability density ${\left| {\Psi \left( {x,t} \right)} \right|^2}$ is coded with color. Red represent high value, and gray represent $0$. (a) Plot of the quantum carpet from $t=0$ to $t=T$. Note the reconstruction of wave function ${\Psi \left( {x,t} \right)}$ at $t=T$. (b) and (c) show the details of the quantum carpet at $t<0.1T$ and $t>0.9T$. (d) A mirror image of the initial state formed at $t=T/2$.} \label{l008}
		\end{figure}

			We consider a particle of mass $m$ trapped in an infinite square well with the potential defined by
			\begin{equation}
				V\left( x \right) = \left\{ {\begin{array}{*{20}{c}}0&{x \in \left( {0,L} \right)}\\\infty &{x \notin \left( {0,L} \right)}\end{array}} \right..
			\end{equation}
			The dynamics of the wave function is given by $\Psi \left( {x,t} \right) = \sum\nolimits_{n = 1}^\infty{{c_n}{\psi _n}\left( x \right){e^{ - i{E_n}t/\hbar }}}$. The eigenstates are
			\begin{equation}
				{\psi _n}\left( x \right) = {\eta _x}\sqrt {\tfrac{2}{L}} \sin \left( {\pi nx/L} \right) \ ,
			\end{equation}
			where $\eta_x=1$ if $x \in (0,L)$, otherwise $\eta_x = 0$. At the initial time, we shall focus on a Gaussian wave packet
			\begin{equation} \label{l051}
				\Psi \left( {x,0} \right) = \frac{1}{{\sqrt {\sqrt {2\pi } {s _x}} }}\exp \left[ { - \frac{{{{\left( {x -\bar x} \right)}^2}}}{{4s _x^2}}} \right]\exp \left({i\bar px/\hbar } \right)
			\end{equation}
			as the initial state, where $\bar x$ and $\bar p$ are the expectation values of the position and momentum operator, respectively, and $s_x$ is the standard deviation of the particle position. Recall that the energy eigenvalues are ${E_n} = {n^2}{\pi ^2}{\hbar ^2}/2m{L^2}$, we therefore have
			\begin{equation}\label{l009}
				\Psi \left( {x,t} \right) = \sum\limits_{n = 1}^\infty{{c_n}{\psi _n}\left( x \right)\exp \left( { - i2\pi {n^2}\frac{t}{{4m{L^2}/\pi \hbar }}} \right)},
			\end{equation}
			which means that after the period $t=T$ with
			\begin{equation} \label{l010}
				T = 4m{L^2}/\pi \hbar,
			\end{equation}
			the wave function repeats itself as a ``full revival", i.e., $\Psi \left( {x,T} \right) = \Psi \left( {x,0} \right)$, as shown in Fig. \ref{l008}(a). The details of the evolution for the quantum carpets at $t<0.1T$ and $t>0.9T$ are shown in Fig. \ref{l008}(b) and \ref{l008}(c).
		\par
			Besides full revival, there is also a mirror revival for the wave function at $t=T/2$, i.e., $\Psi \left( {x,T/2} \right) =  - \Psi \left( {L - x,0}\right)$. As shown in Fig. \ref{l008}(d), the wave packet reconstructs itself at $t=T/2$ as a mirror image of the initial packet. This enables to reverse the flow of time and recreate the original impulsive event after time evolution \cite{linkang2016}.
	\subsection{Fractional revival}
		\par
\begin{figure}[htbp]
			\centerline{\includegraphics[width=\linewidth]{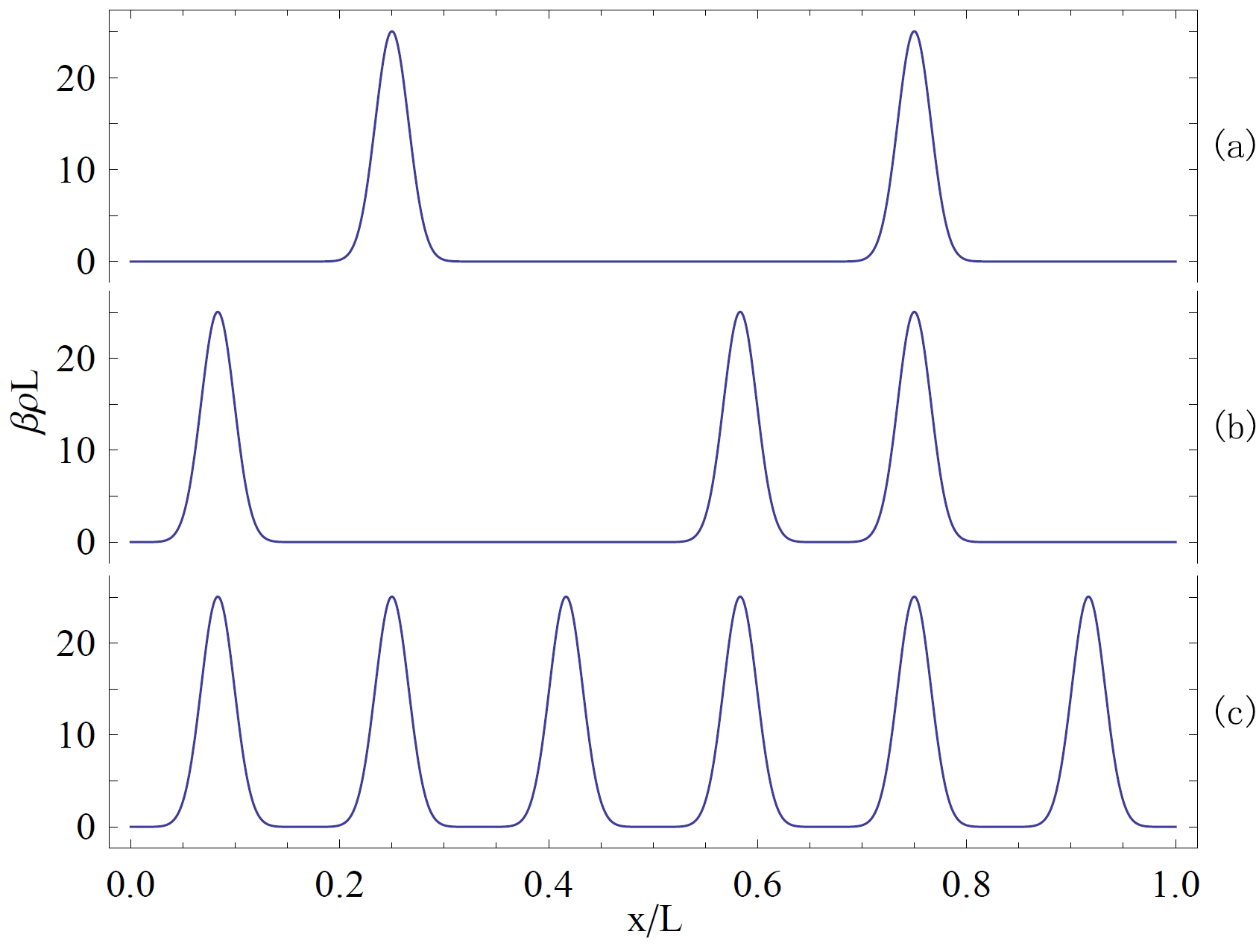}}
			\caption{Fractional revival at $\alpha /\beta=1/2,1/3,5/6$ ($\alpha=1,1,5$ and $\beta=2,3,6$) for (a), (b), and (c), respectively. $x$ is the position of the particle, $L$ is the width of the well, and $\rho={\left| {\Psi \left( {x,t} \right)} \right|^2}$ is the probability distribution where $t=\left( {\alpha /\beta} \right)\left( {T/2} \right)$ and $T$ is the revival time. Parameters $\bar x / L = 1/4$, $s_x/L=1/\left(20\pi\right)$, $\bar p/\left(\hbar/L\right)=25\pi$. We can find that the number of wave packets is identical to the value of $\beta$.} \label{l014}
		\end{figure}
			Apart from the full revival at $t=T$ and the mirror revival at $t=T/2$, there is also fractional revival \cite{naqvi2001fractional,aronstein1997fractional,bernard2017graph} occurring at times $t = \left({\alpha /\beta } \right)\left( {T/2} \right)$, where $\alpha$ and $\beta$ are coprime integers. As we shall see, the wave function
			\begin{equation} \label{l026}
				\Psi \left( {x,\tfrac{\alpha }{\beta }\tfrac{T}{2}} \right) = {\eta _x}\sqrt {\tfrac{2}{L}} \sum\limits_{n = 1}^\infty  {{c_n}\sin \left( {\pi nx/L} \right){e^{ - i\pi {n^2}\alpha /\beta }}}
			\end{equation}
			can be regarded as a superposition of initial states $\Psi \left( {x,0} \right)$ for $\alpha /\beta=1/2,1/3,5/6$, as shown in Fig.~\ref{l014}. Interestingly, one can see that while the number of wave packets is determined by the value of $\beta$, the $\alpha$ affects the position of the wave packets.

\par
			In fact, the same phenomenon of fractional revival was found in the Talbot carpet of classical optics where the wavefront of light reconstructs itself after a grating as a superposition of several initial wavefront. it was well explained with the ``Gaussian sum theory'' \cite{berry2001quantum}. Inspired by this, the new formula of fractional carpet, containing the revival amplitudes and absolute phases at the same time, is given with the help of Gaussian sum.

To be more accurate, the wave function is a superposition of the odd-extended initial state $\Phi(x,0)$, where the odd extension means
			\begin{equation}\label{l035}
				\Phi (x,t) = \sum\limits_{n =  - \infty }^\infty  {\tilde \Psi \left( {x - 2Ln,t} \right)},
			\end{equation}
			and
			\begin{equation}\label{l101}
				\tilde\Psi(x,t)=\Psi(x,t)-\Psi(-x,t).
			\end{equation}
			There are $\beta$ copies of $\Phi(x,0)$ in fractional revival, and each of the copies contributes only one initial packet $\Psi \left( {x,0} \right)$ to the revival.

		\par
			To form fractional revival, we can write the expansion coefficient by
			\begin{equation}
				{c_n} = \sqrt {2/L} \int_0^L {\sin \left( {\pi nx/L} \right)\tilde \Psi \left( {x,0} \right)dx} \ .
			\end{equation}
			Using the identity $\sin \theta  = \left( {{e^{i\theta }} - {e^{ - i\theta }}} \right)/2i$, it can be rewritten by
			\begin{equation} \label{l023}
				{c_n} = \tfrac{1}{{2i}}\sqrt {\tfrac{2}{L}} \int_{ - \infty }^\infty  {{e^{i\pi nx/L}}\tilde \Psi \left( {x,0} \right)dx}.
			\end{equation}
			Now, let us extend the definition of $c_n$ for $n=0,\pm1,\pm2,\pm3,...$ such that $c_{-n}=-c_n$, and hence
			\begin{equation} \label{l028}
				\Psi \left( {x,\tfrac{\alpha }{\beta }\tfrac{T}{2}} \right) = \frac{{{\eta _x}}}{{2i}}\sqrt {\frac{2}{L}} \sum\limits_{n =  - \infty }^\infty  {{c_n}{e^{i\pi nx/L}}{e^{ - i\pi {n^2}\alpha /\beta }}}.
			\end{equation}
			Substituting Eq. (\ref{l023}) into Eq. (\ref{l028}) and switching the order between summation and integration, we get
			\begin{equation} \label{l030}
				\Psi \left( {x,t} \right) = \frac{{{\eta _x}}}{{2L}}\int_{ - \infty }^\infty  {\tilde \Psi \left( {x',0} \right) \times \left[ {\sum\limits_{n =  - \infty }^\infty  {f\left( n \right)} } \right]} dx' \ ,
			\end{equation}
			where $f\left( n \right) \equiv \exp \left( {i\pi n\frac{{x - x'}}{L}} \right)\exp \left( { - i\pi {n^2}\frac{\alpha }{\beta }} \right)$. Introducing the symbol $q_\alpha$, where $q_\alpha=1$ if $\alpha$ is odd and ${q _\alpha } = 0$ if $\alpha$ is even, this expression can be simplified as (see details in Appendix~\ref{2001})
			\begin{equation} \label{l033}
				\Psi \left( {x,\frac{\alpha }{\beta }\frac{T}{2}}\right) = \frac{\eta_x}{\beta }\sum\limits_{n =  - \infty }^\infty  {} \tilde \Psi \left( {x - L{q_\alpha } - 2Ln/\beta ,0}\right)S\left( {n,\alpha ,\beta }\right),
			\end{equation}
			where $S\left( {n,\alpha ,\beta } \right) = \sum\nolimits_{j = 1}^\beta  {\exp \left[ {i\pi j\left( {{q_\alpha } + 2n/\beta  - j\alpha /\beta } \right)} \right]}$ is called as the ``Gaussian sum'' \cite{berry2001quantum}. In number theory~\cite{apostol2013introduction}, for any integer $n$ and any coprime integer $\alpha$ and $\beta$, we have
			\begin{equation}
				\left| {S\left( {n,\alpha ,\beta } \right)} \right| =\sqrt \beta \ ,
			\end{equation}
			which implies that
			\begin{equation} \label{l034}
				S\left( {n,\alpha ,\beta } \right) = \sqrt \beta  \exp\left[ {i\Theta \left( {n,\alpha ,\beta } \right)}\right],
			\end{equation}
			where $\Theta \left( {n,\alpha ,\beta } \right) = \arg S\left( {n,\alpha ,\beta } \right)$. Substituting Eq. (\ref{l034}) into Eq. (\ref{l033}), and note that $\Theta \left( {\beta n + j,\alpha ,\beta } \right) =\Theta \left( {j,\alpha ,\beta } \right)$, which can be derived easily from its definition, we have
			\begin{equation}
			\begin{split}
				\Psi \left( {x,\frac{\alpha }{\beta }\frac{T}{2}}\right)&= \frac{\eta_x}{{\sqrt \beta }}\sum\limits_{n =  - \infty}^\infty  {\sum\limits_{j = 1}^\beta  {} }\\
				&\tilde \Psi \left( {x - L{q_\alpha } - 2Lj/\beta  - 2Ln,0} \right)\exp \left[ {i\Theta \left( {j,\alpha ,\beta }\right)} \right].
			\end{split}
			\end{equation}
			By switching the order between summations, we finally get
			\begin{equation}
			\begin{split}
				\Psi \left( {x,\frac{\alpha }{\beta }\frac{T}{2}}\right) =&\frac{\eta_x}{{\sqrt \beta  }}\sum\limits_{n = 1}^\beta{}\\
				&\Phi \left( {x - L{q_\alpha } - 2Ln/\beta ,0} \right)\exp \left[ {i\Theta \left( {n,\alpha ,\beta } \right)}\right] \ ,
			\end{split}
			\end{equation}
			where ${\Theta \left( {n,\alpha ,\beta }\right)}$ is absolute phase and $1/\sqrt \beta$ is the equivalent amplitude, which are analytically given at the same time. The relative phase can be obtained by the difference between absolute phases, but not the opposite. It also indicates that, for any initial state $\Psi(x,0)$, the wave function $\Psi\left( {x,t} \right)$ to show the fractional revival at time $t = \left( {\alpha /\beta }\right)\left( {T/2} \right)$ is a superposition of $\beta$ copies of the odd-extended initial wave packet $\Phi \left( x , 0\right)$, separated by $2L/\beta$ in $x$ coordinate.
		\par
\begin{figure}[tbp]
			\centerline{\includegraphics[width=\linewidth]{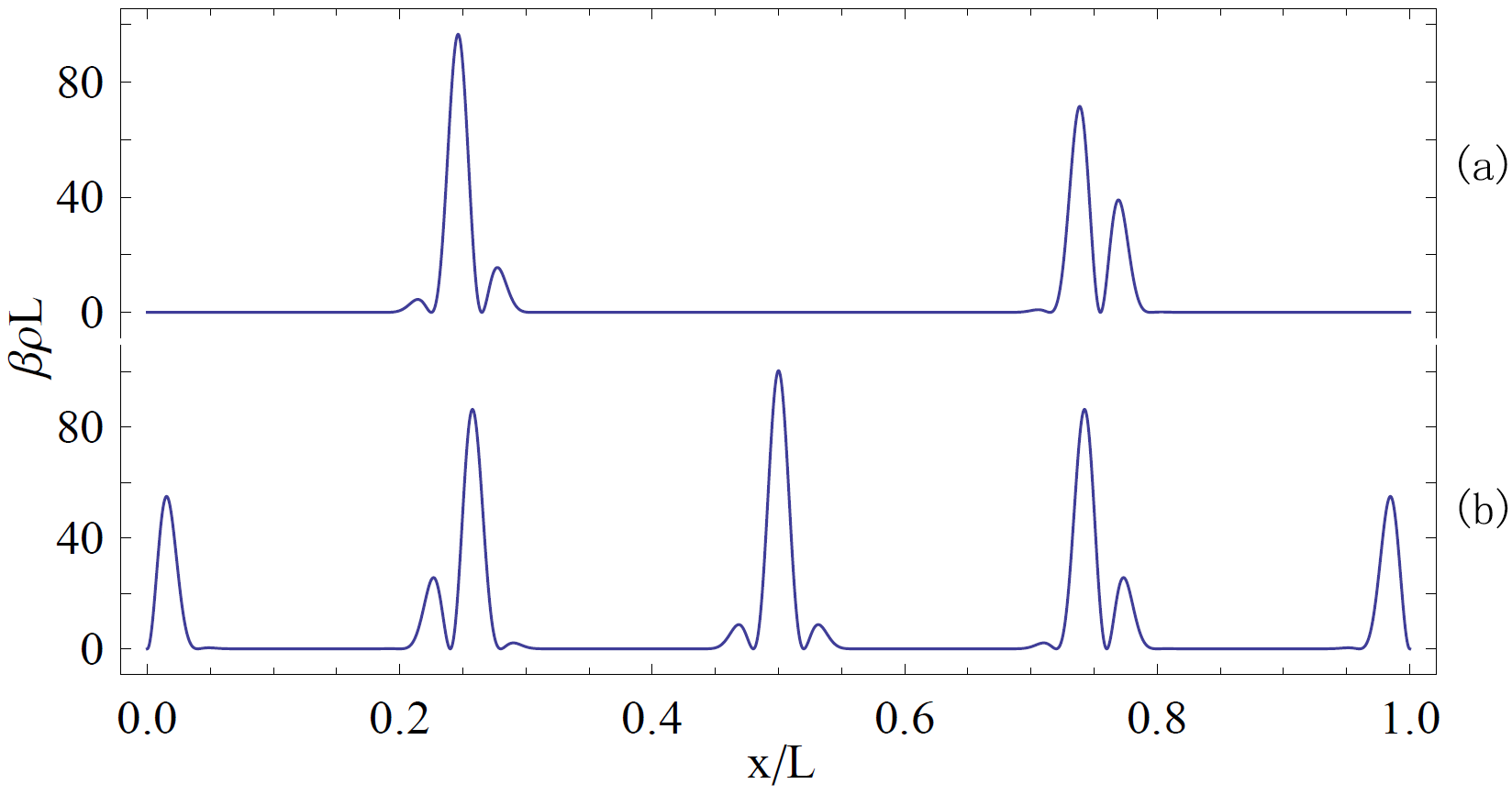}}
			\caption{Fractional revival at $\alpha /\beta=3/4,3/8$ for (a) and (b). $x$	is the position of the particle, $L$ is the width of the well, $\rho={\left| {\Psi \left( {x,t} \right)}\right|^2}$ is the probability distribution where $t=\left( {\alpha /\beta} \right)\left( {T/2} \right)$ and $T$ is the revival time. Parameters $\bar x$, $s_x$, $\bar p$ are the same with Fig. \ref{l014}. There are interference patterns because the superposed wave functions are overlapped.} \label{l017}
		\end{figure}

			If the copies of the extended initial wave packet $\Phi\left( x ,0\right)$ have no overlap with each other, the fractional revival looks like separate wave packets, as shown in Fig. \ref{l014}. If they are overlapped, there will be interference patterns, as shown in Fig. \ref{l017}. The maximum number of initial wave packets we can get in the fractional revival is therefore limited by the width of the packet.
	\subsection{Diagonal canals}
\begin{figure}[tbp]\centerline{\includegraphics[width=\linewidth]{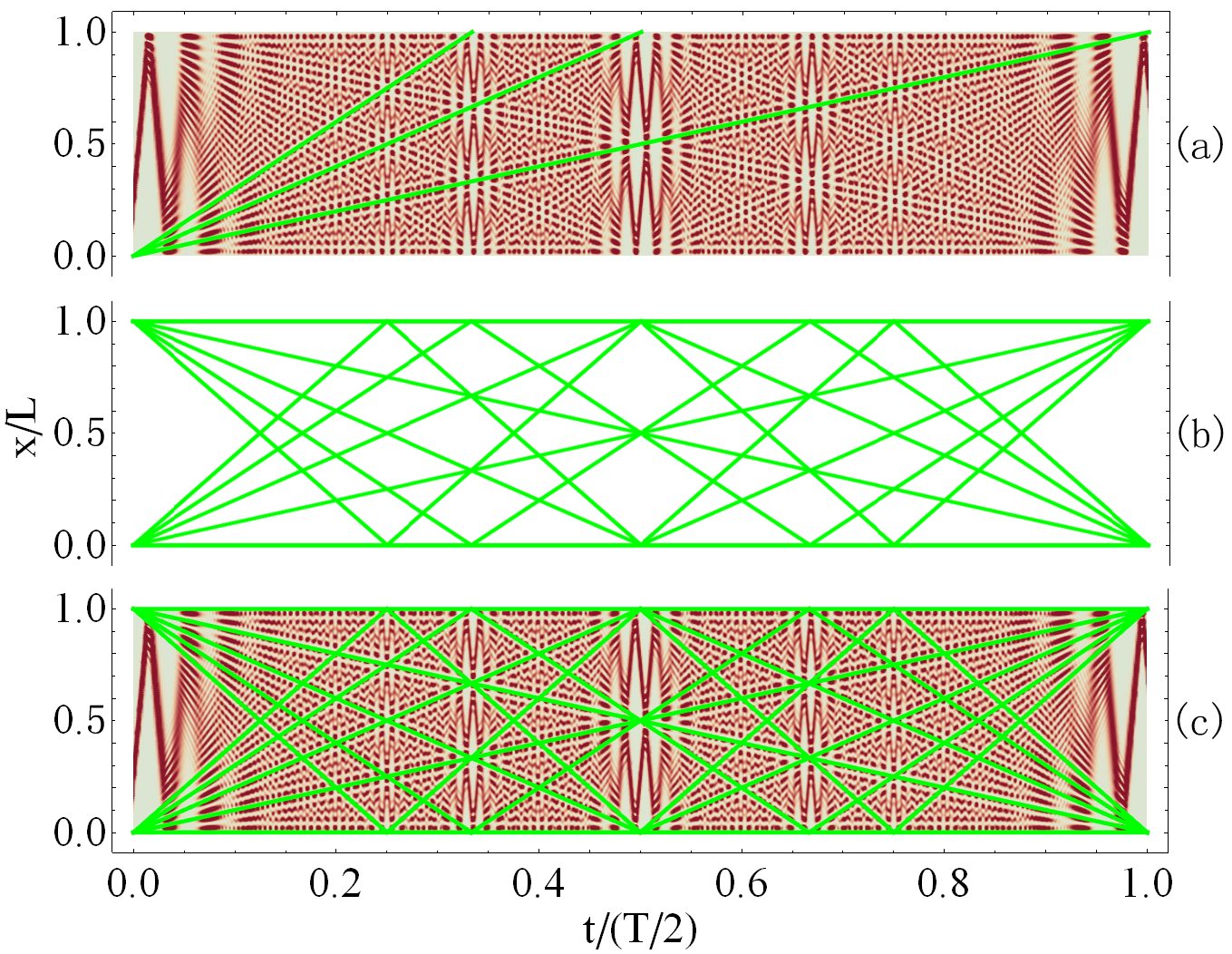}}
			\caption{Diagonal canals in the quantum carpet. The color coding and parameters $\bar x$, $s_x$, $\bar p$ are the same with Fig. \ref{l008}. (a) 3 diagonal canals of the quantum carpet emphasized by green lines. (b) Straight lines that satisfy $\tilde x=0$. Integer $j$ is from $-4$ to $4$ and integer $k$ is from $- \left| j \right|$ to $\left| j\right| + 1$. (c) A combination of (a) and (b) shows that every green line is the center of a group of parallel canals.} \label{l039}
		\end{figure}
		\par
			The diagonal canal of quantum carpet refers to the groups of gray lines that traverse the carpet from side to side, as shown in the green lines of Fig. \ref{l039}(a). This phenomenon can be explained by decomposing the carpet into background and interference terms with the help of Wigner function ${W_f}\left( {x,p,t} \right)$ \cite{stifter1997teilchen}. It is defined as
			\begin{equation} \label{l041}
			\begin{split}
				{W_f} = \frac{1}{{\pi \hbar }}\int_{ - \infty }^\infty  {{f^*}\left( {x + y,t} \right)f\left( {x - y,t} \right){e^{2ipy/\hbar }}dy},
			\end{split}
			\end{equation}
			where $f(x,t)$ is an arbitrary function of variables $x$ and $t$.
		\par
			From Eq. (\ref{l041}) and Eq. (\ref{l035}), we get
			\begin{equation} \label{l0321}
			\begin{split}
			{W_\Phi }\left( {x,p,t} \right) &=\frac{1}{{\pi \hbar }}\sum\limits_{j,k =  - \infty }^\infty  {\int_{ - \infty }^\infty  {} } \\
			&{{\tilde \Psi }^*}\left( {x + y - 2Lj,t} \right)\tilde\Psi \left( {x - y - 2Lk,t} \right){e^{2ipy/\hbar }}dy.
			\end{split}
			\end{equation}
			To simplify the notations, define variables $\tilde x$ and $\tilde p$ as
			\begin{equation} \label{l068}
				\begin{array}{*{20}{c}}{\tilde x = \left( {\frac{x}{L} - j\frac{t}{{T/2}} - k} \right)L,}&{\tilde p = \frac{{\pi \hbar }}{{2L}}j}\end{array},
			\end{equation}
			where $j$ and $k$ are integers. We will see later that every straight line in the $x-t$ plane satisfying $\tilde x=0$ is the center of a group of parallel canals, as shown in Fig. \ref{l039}(b) and \ref{l039}(c). After several mathematical transformation (see Appendix \ref{l057} for details), Eq. (\ref{l0321}) can by modified as
			\begin{equation}
			\begin{split}
			{W_\Phi }\left( {x,p,t} \right) =& \frac{{\pi \hbar }}{{2L}}\sum\limits_{j,k =  - \infty }^\infty  {} \\
			&{\left( { - 1} \right)^{jk}}{W_{\tilde \Psi }}\left({x - Lk,p,t} \right)\delta \left( {p - \tilde p} \right).
			\end{split}
			\end{equation}
		\par
			There are two useful properties of Wigner function, which can be easily derived from its definition. First, ${\left| f \right|^2}$ can be obtained by ${\left| f \right|^2} = \int_{ - \infty }^\infty  {{W_f}dp}$. Second, if the given function $f$ satisfies the Schr\"{o}dinger equation of a free particle $i\hbar \frac{\partial }{{\partial t}}f =  - \frac{{{\hbar ^2}}}{{2m}}\frac{{{\partial ^2}}}{{\partial {x^2}}}f$ and the boundary condition $f\left( { \pm \infty ,t} \right) = 0$, then the time evolution of Wigner function can be written by ${W_f}\left( {x,p,t} \right) = {W_f}\left( {x - \frac{p}{m}t,p,0} \right)$. It is obvious that ${\left| {\Psi \left( {x,t} \right)} \right|^2} = \eta_x{\left|{\Phi \left( {x,t} \right)} \right|^2}$, implying that ${\left| {\Psi \left( {x,t} \right)} \right|^2} = \eta_x\int_{ -\infty }^\infty  {{W_\Phi }\left( {x,p,t} \right)dp}$. This gives
			\begin{equation}
				{\left| {\Psi \left( {x,t} \right)} \right|^2} = {\eta _x}\frac{{\pi \hbar }}{{2L}}\sum\limits_{j,k =  - \infty }^\infty  {{{\left( { - 1} \right)}^{jk}}{W_{\tilde \Psi }}\left( {x - Lk,\tilde p,t} \right)}.
			\end{equation}
			With the help of the second property of Wigner function, we have
			\begin{equation}
				{\left| {\Psi \left( {x,t} \right)} \right|^2} = {\eta _x}\frac{{\pi \hbar }}{{2L}}\sum\limits_{j,k =  - \infty }^\infty  {{{\left( { - 1} \right)}^{jk}}{W_{\tilde \Psi }}\left( {\tilde x,\tilde p,0} \right)}.
			\end{equation}
			By substituting Eq. (\ref{l101}) into the above equation, the probability density ${\left| {\Psi \left( {x,t} \right)} \right|^2}$ for characterizing the quantum carpet can be further decomposed into background and interference terms as \cite{stifter1997teilchen}
			\begin{equation}
				{\left| {\Psi \left( {x,t} \right)} \right|^2} =\eta_x \frac{{\pi \hbar }}{{2L}}\sum\limits_{j,k =  - \infty }^\infty{\left( {B_{j,k}^ +  + B_{j,k}^ -  + {I_{j,k}}} \right)}.
			\end{equation}
			It can always be achieved whatever the initial state $\Psi(x,0)$ is. The background terms $B_{j,k}^\pm$ can be expressed by the Wigner function of the initial state $\Psi \left( {x,0}\right)$ as $B_{j,k}^ \pm  = {\left( { - 1} \right)^{jk}}{W_\Psi }\left( { \pm \tilde x, \pm \tilde p,0} \right)$. And the interference terms ${I_{j,k}} = {\left( { - 1} \right)^{jk}}I\left( {\tilde x,\tilde p,0}\right)$ can be obtained from $I(x,p,t) = {W_{\tilde \Psi }}(x,p,t) - {W_\Psi }(x,p,t) - {W_\Psi }(-x,-p,t)$.
		\par
			In our calculations, the initial state $\Psi\left( {x,0} \right)$ is of the form as Eq. (\ref{l051}), so that the specific expressions for $B_{j,k}^\pm$ and $I_{j,k}$ are
			\begin{equation} \label{l052}
				B_{j,k}^ \pm  = \frac{{{{\left( { - 1} \right)}^{jk}}}}{{\pi \hbar }}G\left( {\frac{{ \pm \tilde x - \bar x}}{{{s_x}}}} \right)G\left( {\frac{{ \pm \tilde p - \bar p}}{{{s_p}}}} \right)
			\end{equation}
			and
			\begin{equation}\label{l053}
				{I_{j,k}} =  - 2\frac{{{{\left( { - 1} \right)}^{jk}}}}{{\pi \hbar }}G\left( {\frac{{\tilde x}}{{{s_x}}}} \right)G\left( {\frac{{\tilde p}}{{{s_p}}}} \right)\cos \left( {\frac{{\bar x\tilde p - \bar p\tilde x}}{{\hbar /2}}} \right),
			\end{equation}
			respectively, where $G$ is defined as $G\left( \theta  \right) = \exp \left( { - {\theta ^2}/2}\right)$. The oscillation of interference terms $I_{j,k}$ comes from the ``$\cos$'' factor in Eq. (\ref{l053}), which depends on the variables $\tilde x$ and $\tilde p$.
\begin{figure}[htbp]
			\centerline{\includegraphics[width=\linewidth]{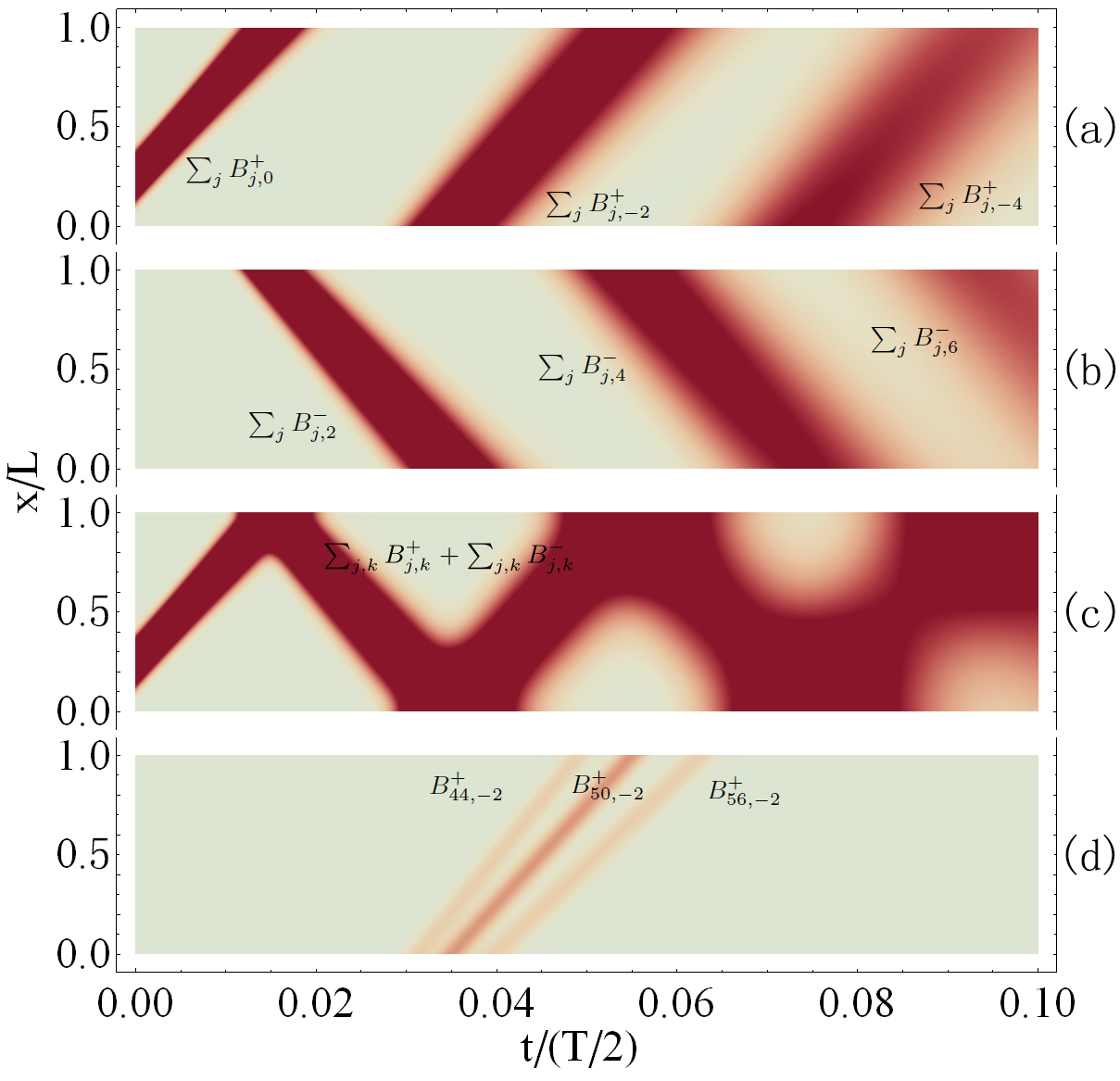}}{}
			\caption{Geometrical correspondences of the background teems. The color coding and parameters $\bar x$, $s_x$, $\bar p$ are the same with Fig. \ref{l008}. (a) ${B_{j,k}^ +}$ propagate in the direction of $\bar p$. $\sum\nolimits_j {B_{j,0}^ + }$, $\sum\nolimits_j {B_{j,-2}^ + }$, and $\sum\nolimits_j {B_{j,-4}^ + }$ are shown from left to right. (b) ${B_{j,k}^ - }$ propagate against the direction of $\bar p$. $\sum\nolimits_j {B_{j,2}^ - }$, $\sum\nolimits_j {B_{j,4}^ - }$, and $\sum\nolimits_j {B_{j,6}^ - }$ are shown from left to right. (c) Total background obtained by $\sum\nolimits_{j,k} {B_{j,k}^ + }  +\sum\nolimits_{j,k} {B_{j,k}^ - }$. (d) The summation over $j$ leads to the diffusion of wave packet. ${B_{44, - 2}^ + }$, ${B_{50, - 2}^ + }$ and	${B_{56, - 2}^ + }$ are shown from left to right.} \label{l072}
		\end{figure}
 \par
\begin{figure}[htbp]
			\centerline{\includegraphics[width=\linewidth]{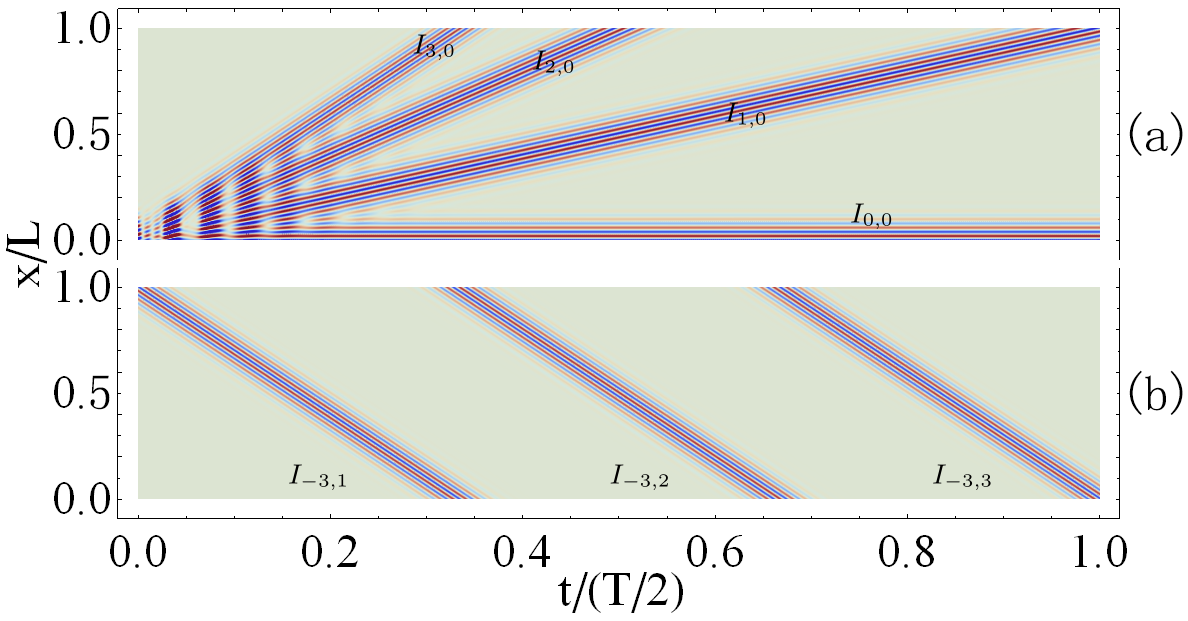}}
			\caption{Geometrical interpretation of the interference teems. The color coding and parameters $\bar x$, $s_x$, $\bar p$ are the same with Fig \ref{l008}. In addition, negative value is represented by blue. (a) The slope of interference term $I_{j,k}$ depends on integer $j$. In this subfigure, $I_{0,0}$, $I_{1,0}$, $I_{2,0}$, and $I_{3,0}$ are shown from bottom to top. All of them are centered along straight line $\tilde x=0$ with $j=0,1,2,3$ and $k=0$. (b) The intercept of interference term $I_{j,k}$ depends on integer $k$. In this subfigure, $I_{-3,1}$, $I_{-3,2}$, and $I_{-3,3}$ are shown from left to right. All of them are centered along straight line $\tilde x=0$ with $j=-3$ and $k=1,2,3$.} \label{l073}
		\end{figure}

 The analysis about the relations between background/interference terms and their geometrical correspondences are given in Fig. \ref{l072} and \ref{l073} with notation $\sum\nolimits_\theta {}  =\sum\nolimits_{\theta =  - \infty }^\infty  {} $. ${B_{j,k}^ +}$ propagates in the direction of $\bar p$ while ${B_{j,k}^ - }$ propagate against, as shown in Fig. \ref{l072}(a) and \ref{l072}(b). The total background can be obtained by $\sum\nolimits_{j,k} {B_{j,k}^ + }  +\sum\nolimits_{j,k} {B_{j,k}^ - }$, as shown in Fig. \ref{l072}(c). In fact, only the terms $\sum\nolimits_j {B_{j,k}^ \pm }$ with even $k$ will contribute to the total background. The summation $\sum\nolimits_j {B_{j,k}^ \pm }$ will oscillate to $0$ if $k$ is odd, because of the factor ${\left( { - 1} \right)^{jk}}$ in $B_{j,k}^\pm$. As shown in Fig. \ref{l072}(d), the summation over $j$ leads to the diffusion of wave packet.
 \par
 Moreover, it is quite clear to get the geometrical interpretations of interference terms. Every $I_{j,k}$ digs several parallel canals along straight line $\tilde x = 0$ with different slope controlled by $j$ (shown in Fig. \ref{l073}(a)) and intercept controlled by $k$ (shown in Fig. \ref{l073}(b)).

\section{discrete quantum carpets} \label{l003}
	\subsection{Theoretical analysis}
		\par
\begin{figure}[htbp]
			\centerline{\includegraphics[width=\linewidth]{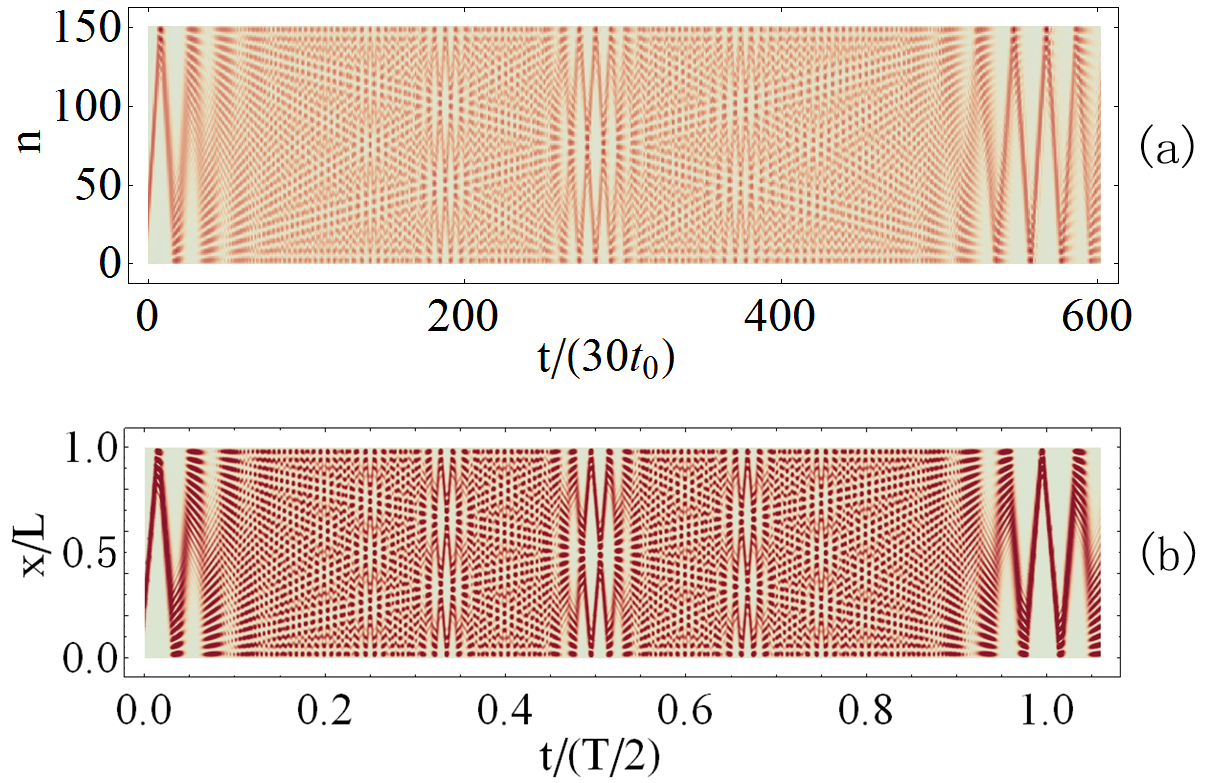}}
			\caption{Discrete quantum carpet (a) compared with the continuous one (b). The color coding and parameters $\bar x$, $s_x$, $\bar p$ are the same with Fig \ref{l008}. $n$ is the label for position eigenstate $\left| n \right\rangle $, $t$ is the evolution time, $t_0$ is a time unit defined by $t_0=\hbar/J$ and $T$ is given by Eq. (\ref{l010}).} \label{l077}
		\end{figure}

			Let $\left| 1 \right\rangle ,\left| 2 \right\rangle ,..., \left| N \right\rangle $ be a set of complete orthogonal basis of a quantum system. Then a Hamiltonian with nearest neighbor interaction takes the form of
			\begin{equation} \label{l100}
				\hat H = -\frac{J}{2}\sum\limits_{n = 1}^{N - 1} {\left({\left| {n + 1} \right\rangle \left\langle n \right| +\left| n 	\right\rangle \left\langle {n + 1} \right|}\right)}
			\end{equation}
			where $J$ represents the coupling strength. This Hamiltonian is equivalent to infinite square well if the basis $\left| 1 \right\rangle ,\left| 2 \right\rangle ,...,\left| N \right\rangle $ are regarded as position eigenstates and their total number $N$ is large enough. The discrete version of the initial state of Eq. (\ref{l051}) is
			\begin{equation} \label{l076}
				\left| {{\Psi _0}} \right\rangle  = A\sum\limits_{n = 1}^N {\exp \left[ { - \frac{{{{\left( {{x_n} - \bar x}\right)}^2}}}{{4s _x^2}}} \right]\exp \left( {i\bar p{x_n}/\hbar } \right)\left| n \right\rangle },
			\end{equation}
			where $A = {\left[ {\sum\nolimits_{n = 1}^N {G\left( {\left( {{x_n} - \bar x} \right)/{s_x}} \right)} } \right]^{ - 1/2}}$ is the normalization coefficient and $x_n=nL/N$ are points equally spaced in the discrete well. By 	solving the dynamical Schr\"{o}dinger equation $i\hbar \frac{\partial }{{\partial t}}\left| \Psi \right \rangle  = \hat H\left| \Psi\right\rangle$ with initial condition in Eq. (\ref{l076}), we get the time-dependent state
			\begin{equation}
				\left| {{\Psi _t}} \right\rangle  = \hat U ^\dag\hat D \hat U \left| {{\Psi _0}} \right\rangle,
			\end{equation}
			where the operator $\hat U$ is unitary with elements
			\begin{equation}
				\left\langle j \right|\hat U\left| k \right\rangle=\sqrt {\tfrac{2}{{N + 1}}} \sin \left( {\pi \tfrac{{jk}}{{N+ 1}}} \right).
			\end{equation}
			The operator $\hat D$ is diagonal with elements
			\begin{equation}
				\left\langle n \right|\hat D\left| n \right\rangle  =\exp \left( { - i{\varepsilon _n}t/\hbar } \right),
			\end{equation}
			where $\varepsilon _n$ are the energy eigenvalues with
			\begin{equation}
				{\varepsilon _n} =- J\cos \left( {\pi \tfrac{n}{{N + 1}}}\right).
			\end{equation}
			A numerical calculation of discrete carpet ${\left| {\left\langle {n} \mathrel{\left | {\vphantom {n {{\Psi _t}}}}\right. \kern-\nulldelimiterspace} {{{\Psi _t}}} \right\rangle } \right|^2}$ where $N=150$ is given in Fig. \ref{l077}(a) with the horizontal coordinate $t$ and vertical coordinate $n$. It shows the same manner with the continuous one (Fig. \ref{l077}(b)).
	\subsection{Experimental instructions}
		\begin{figure}[htbp]
			\centerline{\includegraphics[width=\linewidth]{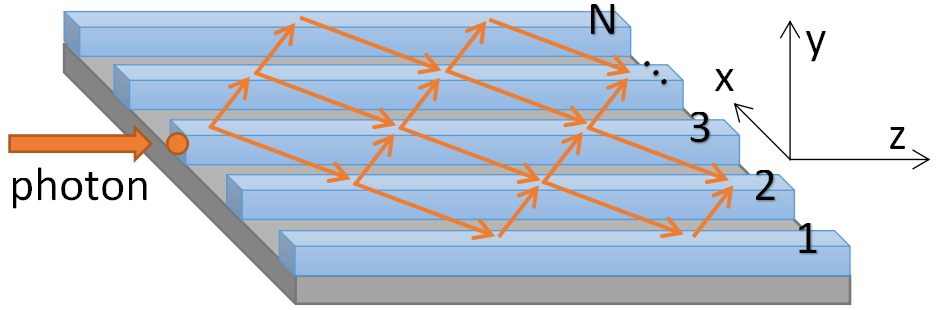}}
			\caption{Waveguide array. A photon can propagate in it and form a carpet in the $x-z$ plane.} \label{l080}
		\end{figure}
		\par
			A feasible system for observing the discrete carpet is the waveguide array, as shown in Fig. \ref{l080}. A photon propagates in a waveguide array with $N$ single-mode channel waveguides laid parallel to each other can be described by position eigenstates $\left| 1 \right\rangle ,\left| 2\right\rangle ,...,\left| N \right\rangle $, where the state $\left| n \right\rangle$ represents that the photon is in the waveguide sited at $x_n=nL/N$. The gap between those waveguides is designed to make their individual modes overlap. In this case, the propagation between waveguides ($x$ direction) satisfies the Hamiltonian in Eq. (\ref{l100}) \cite{christodoulides}.
		\par
			The propagation of photon along waveguides ($z$ direction) is just uniform motion, thus transforms the time evolution into space variation. We can input the same initial state at $z=0$ and detect the photon at $z=z_n$. Through the statistical distribution of photons, the discrete quantum carpet at the moment $t=z_n/c$ will be obtained. Combining the distribution at $z_1,z_2,...,z_N$ together, we will get the discrete quantum carpet in experiment.
		\par
			As shown in Fig. \ref{l077}, $N=150$ is sufficient to observe a clear pattern of carpet while the experiments can couple waveguides up to $10^4$ channels \cite{perets2008realization}, which can fully satisfy the requirement.
	\section{summary} \label{l004}
		In summary, we analyzed the three essential features of continuous (full revival, fractional revival and diagonal canal) and discrete quantum carpets. For the fractional revival, it is interpreted as a superposition of several odd-extended initial wave packets with the same amplitude, and their absolute phases are obtained from the ``Gaussian sum''. The relations between background/interference terms of diagonal canals and their geometric representations are analyzed explicitly. The summation of background terms forms the diffusion, reconstruction, and reflection of wave packet, and each interference term constructs several parallel canals in the carpet. With regard to the discrete carpet, we proposed a method for its experimental observation with current optical waveguide technology. The results will improve the understanding of the relation between fundamental ``particle in box'' model and the intricate quantum carpet patterns.
		
			\section*{Acknowledgements} This work is supported by Natural Science Foundation of Guangdong Province (2017B030308003) and the Guangdong Innovative and Entrepreneurial Research Team Program (No.2016ZT06D348), and the Science Technology and Innovation Commission of Shenzhen Municipality (ZDSYS20170303165926217, JCYJ20170412152620376), and the National Natural Science Foundation of China (No. 61704071), and the Postdoctoral Science Foundation
of China (No.2018M632195).
\begin{appendix}
\section{Decomposition of the summation}\label{2001}
	\par
		Note that we can always divide the summation into groups $\sum\nolimits_{n =  - \infty }^\infty  {f\left( n \right)}  = \sum\nolimits_{n =  - \infty }^\infty  {\sum\nolimits_{j = 1}^\beta  {f\left( {\beta n + j} \right)} }$. For example, if $\beta=3$, it has $...+f(-2)+f(-1)+f(0)+f(1)+f(2)+f(3)+f(4)+f(5)+f(6)+...=...+[f(-2)+f(-1)+f(0)]+[f(1)+f(2)+f(3)]+[f(4)+f(5)+f(6)]+...$. In this way, we have
		\begin{equation}\label{l200}
		\begin{split}
			&{\sum\limits_{n =  - \infty }^\infty  {\exp \left( {i\pi n\frac{{x - x'}}{L}} \right)\exp \left( { - i\pi {n^2}\frac{\alpha }{\beta }} \right)} }=\\
			&\left\{ {\sum\limits_{n =  - \infty }^\infty  {\exp\left[ {i\pi n\beta \left( {\frac{{x - x'}}{L} - {q _\alpha }} \right)} \right]} } \right\} \times\\
			&\left\{ {\sum\limits_{j = 1}^\beta  {\exp \left[ {i\pi j\left( {\frac{{x - x'}}{L} - j\frac{\alpha }{\beta }}\right)} \right]} } \right\},
		\end{split}
		\end{equation}
		where ${q _\alpha }=1$ if $\alpha$ is odd and ${q _\alpha }=0$ if $\alpha$ is even. Employing the Poisson summation formula \cite{deitmar2014principles}
		\begin{equation} \label{l061}
			\sum\limits_{n = - \infty }^\infty  {{e^{i2\pi n\theta }}}  = \sum\limits_{n =  - \infty }^\infty  {\delta \left({n - \theta } \right)},
		\end{equation}
		with $\delta$ being the Dirac delta function, the expression in (\ref{l200}) can be given by
		\begin{equation}\label{l029}
		\begin{split}
			\frac{{2L}}{\beta }\sum\limits_{n =  - \infty }^\infty{\sum\limits_{j = 1}^\beta  {\delta \left( \kappa \right)  } }
			\exp \left[ {i\pi j\left( {{q_\alpha } + \frac{{2n - j\alpha }}{\beta }} \right)} \right],
		\end{split}
		\end{equation}
		where $\kappa={x - x' - L{q_\alpha } - 2Ln/\beta }$. Substituting Eq. (\ref{l029}) into Eq. (\ref{l030}), and switching the order between summation and integration, then Eq. (\ref{l033}) is obtained.
\section{Pattern decomposition} \label{l057}
	\par For the Wigner function being of the form
		\begin{equation}
		\begin{split}
			{W_\Phi }\left( {x,p,t} \right) &=\frac{1}{{\pi \hbar }}\sum\limits_{j,k =  - \infty }^\infty  {\int_{ - \infty }^\infty  {} } \\
			&{{\tilde \Psi }^*}\left( {x + y - 2Lj,t} \right)\tilde\Psi \left( {x - y - 2Lk,t} \right){e^{2ipy/\hbar }}dy,
		\end{split}
		\end{equation}
		replacing $y$ by $y=z+2Lj$ and switching the order in the summation by
		\begin{equation}
			\sum\limits_{j,k =  - \infty }^\infty  {f\left( {j,k}\right)}  = \sum\limits_{j,k =  - \infty }^\infty  {f\left( {j,k - j} \right)},
		\end{equation}
		we will get
		\begin{equation}
		\begin{split}
			{W_\Phi }\left( {x,p,t} \right) =& \frac{1}{{\pi \hbar }}\left( {\sum\limits_{j =  - \infty }^\infty  {{e^{4ipLj/\hbar }}} } \right)\sum\limits_{k =  - \infty }^\infty{\int_{ - \infty }^\infty  {} }\\
			&{{\tilde \Psi }^*}\left( {x + z,t} \right)\tilde \Psi\left( {x - z - 2Lk,t} \right){e^{2ipz/\hbar }}dz.
		\end{split}
		\end{equation}
		Using the Poisson summation formula (\ref{l061}) and replacing $z$ by $z=y-Lk$, we have
		\begin{equation} \label{l062}
		\begin{split}
			{W_\Phi }\left( {x,p,t} \right) =& \frac{{\pi \hbar }}{{2L}}\sum\limits_{j,k =  - \infty }^\infty  {} \\
			&{\left( { - 1} \right)^{jk}}{W_{\tilde \Psi }}\left({x - Lk,p,t} \right)\delta \left( {p - \tilde p} \right).
		\end{split}
		\end{equation}
		 Then it is easy to obtain
		\begin{equation} \label{l067}
			{\left| {\Psi \left( {x,t} \right)} \right|^2} = {\eta _x}\frac{{\pi \hbar }}{{2L}}\sum\limits_{j,k =  - \infty }^\infty  {{{\left( { - 1} \right)}^{jk}}{W_{\tilde \Psi }}\left( {x - Lk,\tilde p,t} \right)}.
		\end{equation}
		From $\tilde \Psi \left( { \pm \infty ,t} \right) = 0$, $i\hbar \frac{\partial }{{\partial t}}\tilde \Psi \left({x,t} \right) =  - \frac{{{\hbar ^2}}}{{2m}}\frac{{{\partial ^2}}}{{\partial {x^2}}}\tilde \Psi \left( {x,t}\right)$, and the Eq. (\ref{l067}), we get
		\begin{equation} \label{l069}
			{\left| {\Psi \left( {x,t} \right)} \right|^2} = {\eta _x}\frac{{\pi \hbar }}{{2L}}\sum\limits_{j,k =  - \infty }^\infty  {{{\left( { - 1} \right)}^{jk}}{W_{\tilde \Psi }}\left( {\tilde x,\tilde p,0} \right)}.
		\end{equation}
		Substituting Eq. (\ref{l101}) into Eq. (\ref{l069}), we have
		\begin{equation} \label{l071}
		\begin{split}
			{\left| {\Psi \left( {x,t} \right)} \right|^2} =& \eta_x\frac{{\pi \hbar }}{{2L}}\sum\limits_{j,k =  - \infty }^\infty{{{\left( { - 1} \right)}^{jk}} \times }\\
			&\left[ {{W_\Psi }\left( {\tilde x,\tilde p,0} \right) + {W_\Psi }\left( { - \tilde x, - \tilde p,0} \right) + I\left( {\tilde x, \tilde p,0} \right)}\right],
		\end{split}
		\end{equation}
		where
		\begin{equation}
		\begin{split}
			I\left( {x,p,t} \right)=& {W_{\tilde \Psi }}\left( {x,p,t} \right) - {W_\Psi }\left( {x,p,t} \right) - {W_\Psi }\left( { - x, - p,t}\right)\\
			=&  - \frac{2}{{\pi \hbar }}{\mathop{\rm Re}\nolimits}\int_{ - \infty }^\infty  {{\Psi ^*}\left( {x + y,t}\right) }\Psi \left( { - x + y,t} \right){e^{2ipy/\hbar }}dy.
		\end{split}
		\end{equation}
		Defining the background and interference terms as $B_{j,k}^ \pm  = {\left( { - 1} \right)^{jk}}{W_\Psi }\left( { \pm \tilde x, \pm \tilde p,0} \right)$ and ${I_{j,k}} = {\left( { - 1} \right)^{jk}}I\left( {\tilde x,\tilde p,0}\right)$, we finally get
		\begin{equation}
			{\left| {\Psi \left( {x,t} \right)} \right|^2} = \eta_x\frac{{\pi \hbar }}{{2L}}\sum\limits_{j,k =  - \infty }^\infty{\left( {B_{j,k}^ +  + B_{j,k}^ -  + {I_{j,k}}} \right)}.
		\end{equation}
\end{appendix}
\bibliographystyle{apsrev4-1}
\bibliography{main}
\end{document}